# Medium-temperature furnace bake of Superconducting Radio-Frequency cavities at IHEP

Feisi He[a,b,c], Weimin Pan [a,b,c, d]*, Peng Sha[a,b,c], Jiyuan Zhai[a,b,c], Zhenghui Mi[a,b,c], Xuwen Dai[a,c], Song Jin[a,b,c], Zhanjun Zhang[a,c], Chao Dong[a,b,c], Baiqi Liu[a,b,c], Hui Zhao[a,c], Rui Ge[a,b,c], Jianbing Zhao[a,c], Zhihui Mu[a,c], Lei Du[a,b,c], Liangrui Sun[a,b,c], Liang Zhang[a,c], Conglai Yang[a,c], Xiaobing Zheng[a,c]

[a]Institute of High Energy Physics, Chinese Academy of Sciences, Beijing 100049, China
[b]Key Laboratory of Particle Acceleration Physics & Technology, Institute of High Energy Physics, Chinese Academy of Sciences, Beijing 100049, China
[c]Center for Superconducting RF and Cryogenics, Institute of High Energy Physics, Chinese Academy of Sciences, Beijing 100049, China
[d]University of Chinese Academy of Sciences, Beijing 100049, China

**Abstract**

Recently, heat treatment between 250 °C and 500 °C has been attempted to improve quality factor (Q) of superconducting radio-frequency cavities at FNAL and KEK. Experiments of such medium temperature (mid-T) bake with furnaces have also been carried out at IHEP. Firstly, over ten 1.3 GHz 1-cell cavities were treated with different temperatures at a small furnace, which all demonstrated improvement of Q and anti-Q-slope phenomenon. The average quality factor has reached $3.6 \times 10^{10}$ when the gradient is 16 MV/m, while the highest Q is $4.9 \times 10^{10}$@16MV/m; the maximum gradients of these 1-cell cavities are between 25.1 and 36.9 MV/m. Then, the recipe of mid-T furnace bake at 300 °C for 3 hours has been applied to six 1.3 GHz 9-cell cavities at a new big furnace, which have all shown higher Q and anti-Q-slope at medium field (16~24 MV/m). The average quality factor has reached $3.8 \times 10^{10}$ when the gradient is 16 MV/m. The maximum gradients of the 9-cell cavities are between 22.7 and 26.5 MV/m.

## 1. Introduction

Nowadays, Superconducting radio-frequency (SRF) cavities have been adopted by many accelerators all over the world [1-5]. In order to reduce the construction cost and operational cost of the cryogenics system, Q value of SRF cavities have been increased gradually through many methods in the recent years. For example, Nitrogen doping has been adopted by linac coherent light source II (LCLS-II) in America, which increased Q to $3 \times 10^{10}$ over 20 MV/m for 1.3 GHz 9-cell cavities [6, 7]. In China, Shanghai High repetition rate XFEL and Extreme light facility (SHINE) has begun construction in April 2018, which also require 600 1.3-GHz 9-cell cavities with high Q. In future, the Circular Electron Positron Collider (CEPC) will use 240 650-MHz 2-cell cavities for collider ring and 96 1.3 GHz 9-cell cavities for booster, which require high Q, too [8]. The 1.3 GHz 9-cell cavity of CEPC will be operated at 2.0 K with Q>$3 \times 10^{10}$ at 24 MV/m for the vertical acceptance test. So research on Nitrogen doping has been carried out at Institute of High Energy Physics, Chinese Academy of Sciences (IHEP, CAS), which achieved good result preliminarily [9].

---

*Corresponding author:
 *Email address*: panwm@ihep.ac.cn (Weimin Pan)

Besides Nitrogen doping, a new heat treatment called mid-T bake was also found to improve Q value of SRF cavities recently, which has been carried out and succeeded with 1.3 GHz 1-cell cavities at Fermi National Accelerator Laboratory (FNAL) [10]. The 1.3 GHz 1-cell cavities evacuated and assembled with flanges were heated and maintained above 300 °C in an oven, which can dissolve oxides in the inner surface of cavity. After Mid-T bake, the cavities received vertical test directly with no other operation. The highest Q has reached $5\times10^{10}$ (at 2.0K), while the gradient was over 20 MV/m. And the Bardeen-Cooper-Schrieffer (BCS) and residual resistance both decreased than the EP baseline results.

KEK modified the mid-T bake by heating two 1.3GHz cavities in a vacuum furnace at 400 °C for 3 hours, which is called mid-T furnace bake. After this process, the inner surface of cavity was exposed to air and received high-pressure rinse (HPR), following with assembly in clean room. Both cavities showed high Q ($>3\times10^{10}$) and slight anti-Q-slope phenomenon in the vertical test.

In this paper, we present research of mid-T furnace bake at IHEP. Firstly, eleven 1.3GHz 1-cell cavities received mid-T furnace bake in a small furnace. Different temperatures have been tried between 250 °C and 400 °C, which all indicated increase of Q during vertical tests. The average Q of these 11 cavities reached $3.6\times10^{10}$ at 16MV/m. Then, a bigger furnace was constructed, which was used for 1.3 GHz 9-cell and 650 MHz cavities. Six 9-cell cavities have received mid-T furnace bake at 300 °C for 3 hours, and the improvement of Q value is very exciting, which is resulted from mid-T furnace bake.

## 2. Mid-T furnace bake of 1.3 GHz 1-cell cavities

Fourteen 1.3 GHz 1-cell cavities made of fine-grain bulk niobium had been Electro-polished (EP) and vertical tested as baseline. Gradient of these cavities all exceeded 40 MV/m in the baseline test, which were qualified to receive mid-T furnace bake.

The mid-T furnace bake of 1.3 GHz 1-cell cavities were carried out at a small furnace with dual-vacuum, as Fig. 1. This furnace is equipped with two vacuum systems, which is different from the one used at KEK.

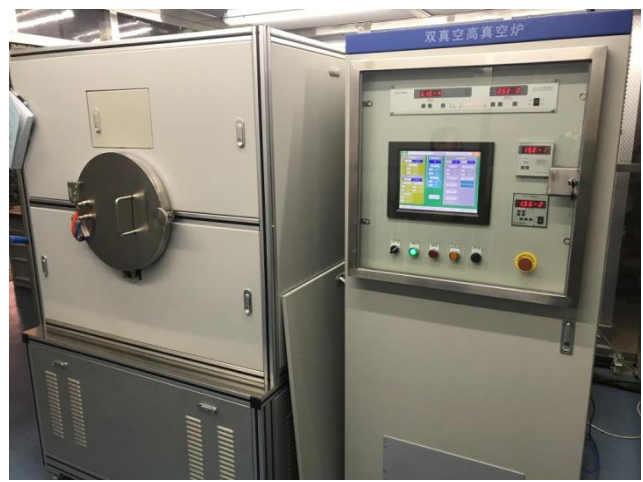

**Fig. 1** Dual-vacuum furnace for 1.3 GHz 1-cell cavity

The small furnace has two chambers: inner chamber and outer chamber. The heater of the furnace is in the outer chamber, which is beneficial to cleanliness of the inner chamber. The inner chamber is made of stainless steel. The 1.3 GHz cavity 1-cell cavity is placed in the inner chamber during mid-T annealing, which is covered with Niobium foil at both flanges, as shown in Fig. 2.

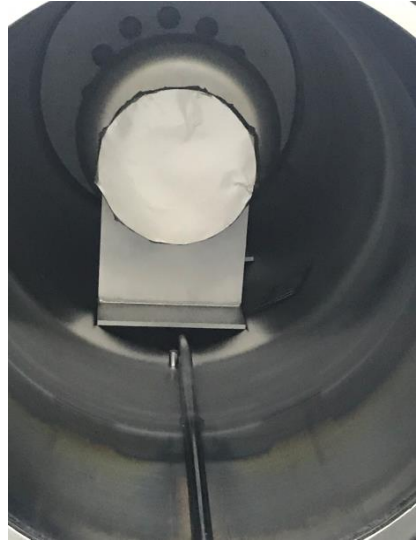

**Fig. 2** Mid-T furnace bake of 1.3 GHz 1-cell cavity

Research of Niobium samples has shown that Oxide layer in the cavity surface can be dissolved by baking above 300 °C [11]. So several recipes of mid-T furnace bake have been attempted with different temperatures: 250 °C, 300 °C, 350 °C and 400 °C, which is for comparison. The annealing time is 3 hours for different temperatures all the time. Firstly, the furnace is evacuated and pumped with a cyropump to achieve high vacuum ($5\times10^{-6}$ Pa). Secondly, the furnace is heated to the medium temperature (250 °C ~ 400 °C) within 2 hours, which maintain for 3 hours. Finally, the cavity begins cooling down naturally to 50 °C and mid-T furnace bake is completed.

The temperature and vacuum during the annealing process is shown as Fig. 3, which takes S25# 1.3 GHz 1-cell cavity for example. High vacuum was maintained with a cryopump during the heating process, and the worst vacuum was $6.8\times10^{-5}$ Pa when the temperature reached 300 °C. After cooling down, the vacuum pressure is better than $1.0\times10^{-6}$ Pa.

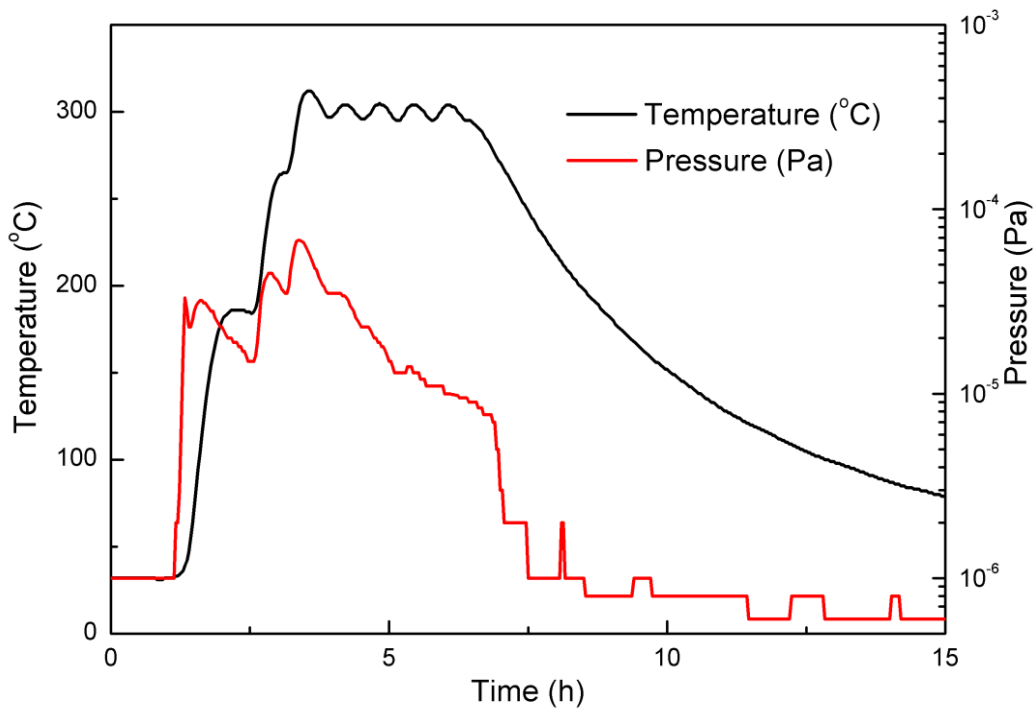

**Fig. 3** Typical process of mid-T furnace bake for S25# with 300 °C. The vacuum was below 1E-4 Pa

during the heat treatment, which decreased to < $1\times10^{-6}$ Pa after cooling down. The other 1.3 GHz 1-cell cavities had similar curves, but with different temperatures (250 °C, 350 °C, 400 °C).

Residual gas analyzer (RGA) measurement was also applied to monitor the composition of residual gas in the furnace (see Fig. 4), which is corresponding to Fig. 3.

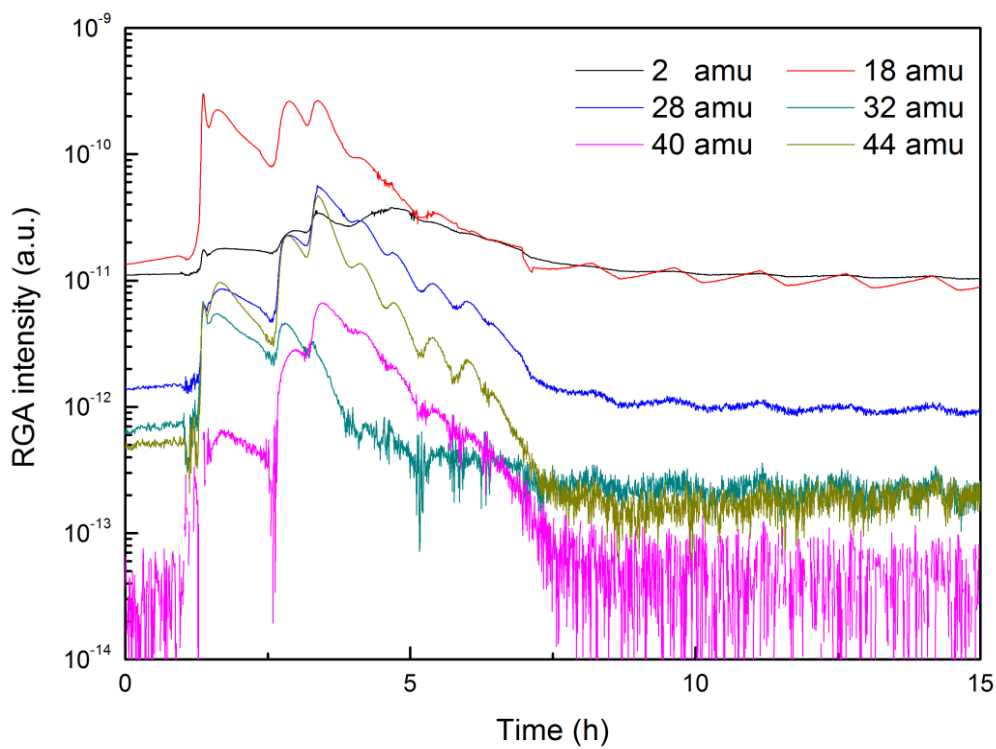

**Fig. 4** Typical RGA curves of mid-T furnace bake for S25# with 300 °C. The lines of different colours at 2, 18, 28, 32, 40 and 44 represent hydrogen, water, nitrogen, oxygen, argon and carbon dioxide respectively.

After mid-T furnace bake, these cavities received high-pressure rinse (HPR) and assembly with flanges in clean room, followed by vertical tests. The 120 °C bake treatment has been cancelled. The vertical test results is shown in Fig. 5. All the cavities indicate high Q in the range of $3\sim5\times10^{10}$ at medium field and exceed the gradient of 25 MV/m. The anti-Q-slope phenomenon appears between 5 ~ 18 MV/m, which is usually observed with Nitrogen doped cavities[12]. The highest Q is $4.9\times10^{10}$@16MV/m, which is achieved by S25#. Average Q of these cavities is $2.9\times10^{10}$ at 24MV/m, which is close to the CEPC spec ($3\times10^{10}$@24MV/m) for vertical test. The result of S17# has reached $1.2\times10^{10}$@37MV/m, which is the highest gradient among all the cavities treated with mid-T furnace bake at IHEP.

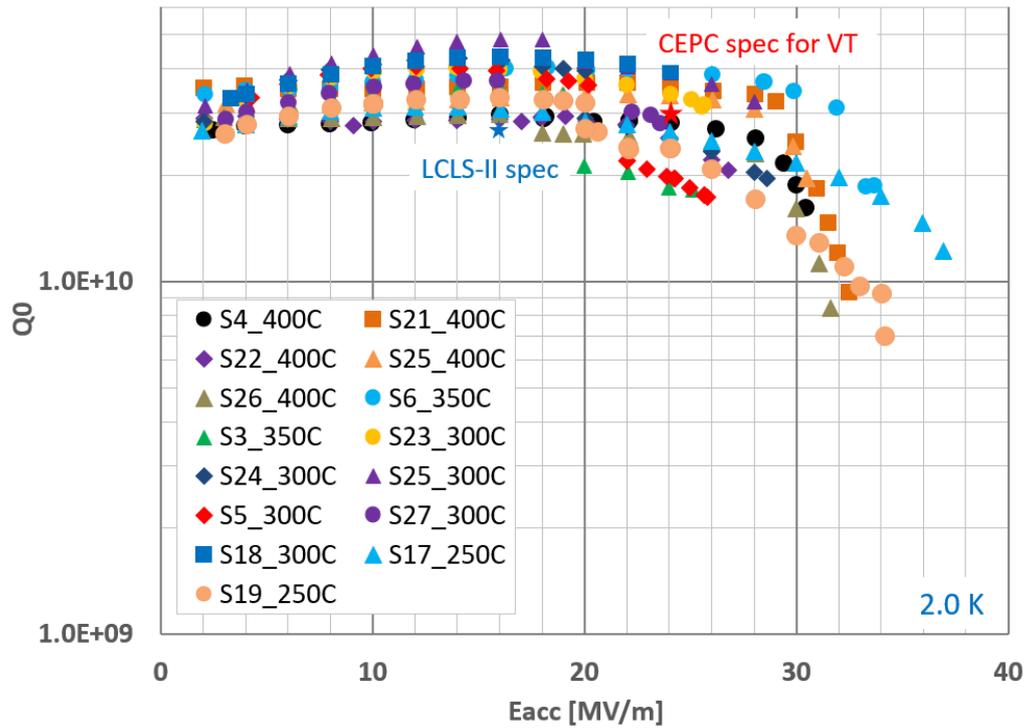

**Fig. 5** Vertical test results of 1.3 GHz 1-cell cavities, which have received mid-T furnace bake with different temperatures

## 3. Mid-T furnace bake of 1.3 GHz 9-cell cavities

According to the successful experience of 1.3 GHz 1-cell cavities above, mid-T furnace bake is appliedy to 1.3 GHz 9-cell cavities, which is carried out at a new big furnace (see Fig. 6). This furnace is equipped with two cryopumps, which helps keeping high vacuum.

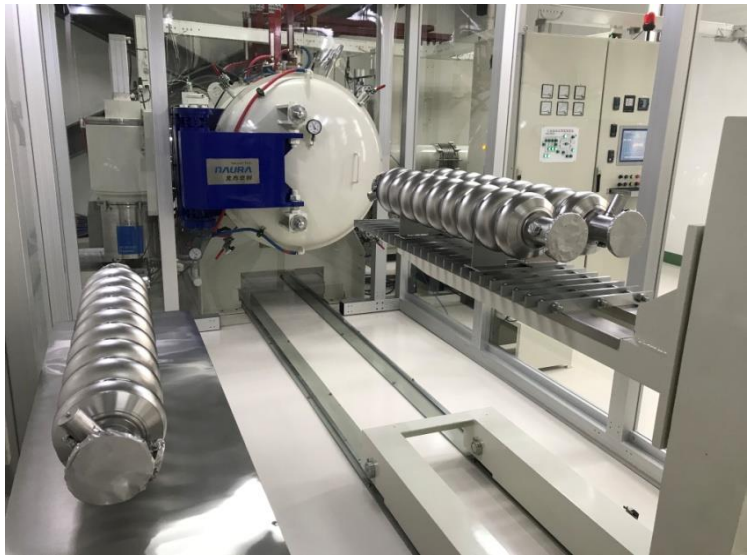

**Fig. 6** Mid-T furnace bake of 1.3 GHz 9-cell cavities at a big furnace

The 300 °C recipe was adopted for the 9cell cavity mid-T furnace bake. Temperature and vacuum during the mid-T furnace bake of two 9-cell cavities (N8# and N9#) are shown in Fig. 7, and the worst vacuum was $2.4\times10^{-5}$ Pa during the heat treatment, which is better than the small furnace. After cooling down, the vacuum pressure was ~$2.8\times10^{-6}$ Pa, which is worse than the small furnace.

Data of RGA during mid-T furnace bake of N8# and N9# has also been recorded as Fig. 8, which is similar as Fig. 4. But the concentration of carbon dioxide is higher than the small furnace. The reason may be that the big furnace is a new one. It is supposed that the composition of carbon dioxide may decrease along with the running of furnace.

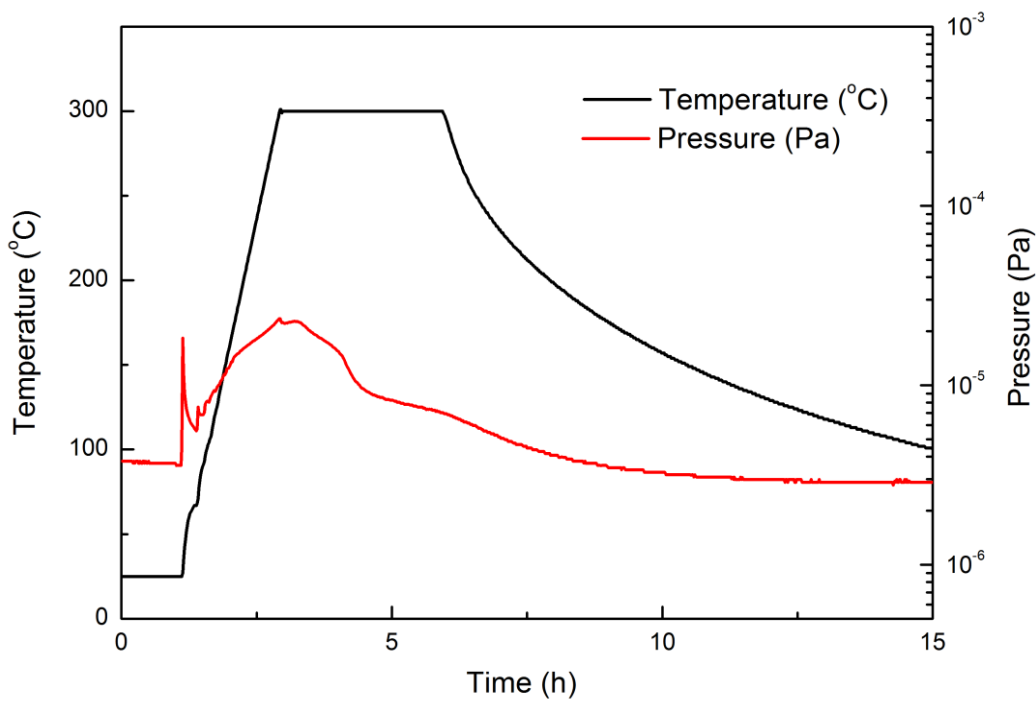

**Fig. 7** Temperature and vacuum variation during mid-T furnace bake of N5# and N7#.

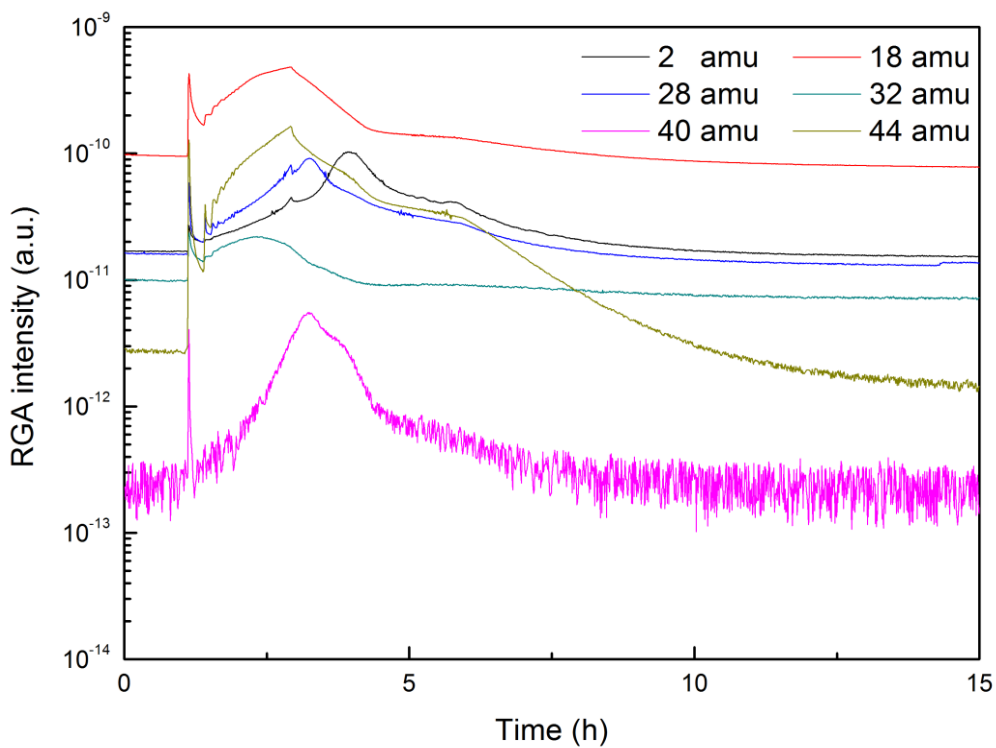

**Fig. 8** RGA curves of mid-T furnace bake for N5# and N7#. The lines of different colours at 2, 18, 28,

32, 40 and 44 represent hydrogen, water, nitrogen, oxygen, argon and carbon dioxide respectively.

Vertical test results of EP baseline and mid-T furnace bake are shown as Fig. 9. Before mid-T furnace bake, five 1.3 GHz 9-cell cavities (N5#, N6#, N7#, N8# and N10#) had all received a baseline vertical test after EP, which all exceeded 30 MV/m. So they're considered to be qualified for following mid-T furnace bake. The highest gradient of 36 MV/m was achieved by N6# among these 5 cavities. After mid-T furnace bake, all the 9-cell cavities demonstrate high Q in the range of $3.5\sim4.4\times10^{10}$ at the gradient between 16~24 MV/m. These cavities have all exceeded the specification of LCLS-II HE ($2.7\times10^{10}$@21MV/m). Five cavities reached the specification of CEPC ($3.0\times10^{10}$@24MV/m), while N9# is exceptional with lower gradient. The average Q of all the six 9-cell cavities is $3.8\times10^{10}$@16MV/m. The highest gradient is 26.5 MV/m, which is achieved by N8#. These results are almost as good as LCLS-II-HE, which have reached $3.5\times10^{10}$@25.7MV/m on average for ten 1.3 GHz 9-cell cavities. The anti-Q-slope behavior of 9-cell cavities is the same as 1-cell cavities.

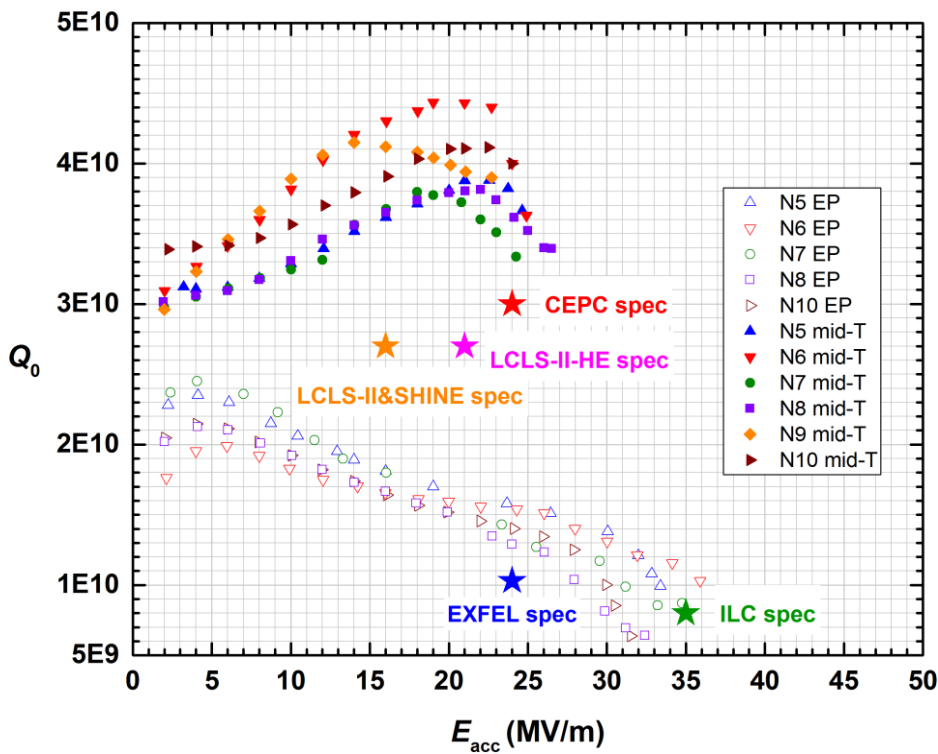

**Fig. 9** Vertical test results of six 1.3 GHz 9-cell cavities (solid icon: mid-T furnace bake; hollow icon: EP baseline)

To analyze the composition of surface resistance, two 9-cell cavities (N8# and N9#) were tested at different temperatures (1.5K ~ 2.0K). The data of Q and gradient at different temperatures is shown as Fig. 10. At 1.5K, Q value of N8# is above $1\times10^{11}$ between 2~20 MV/m, while that of N9# is lower than $1\times10^{11}$.

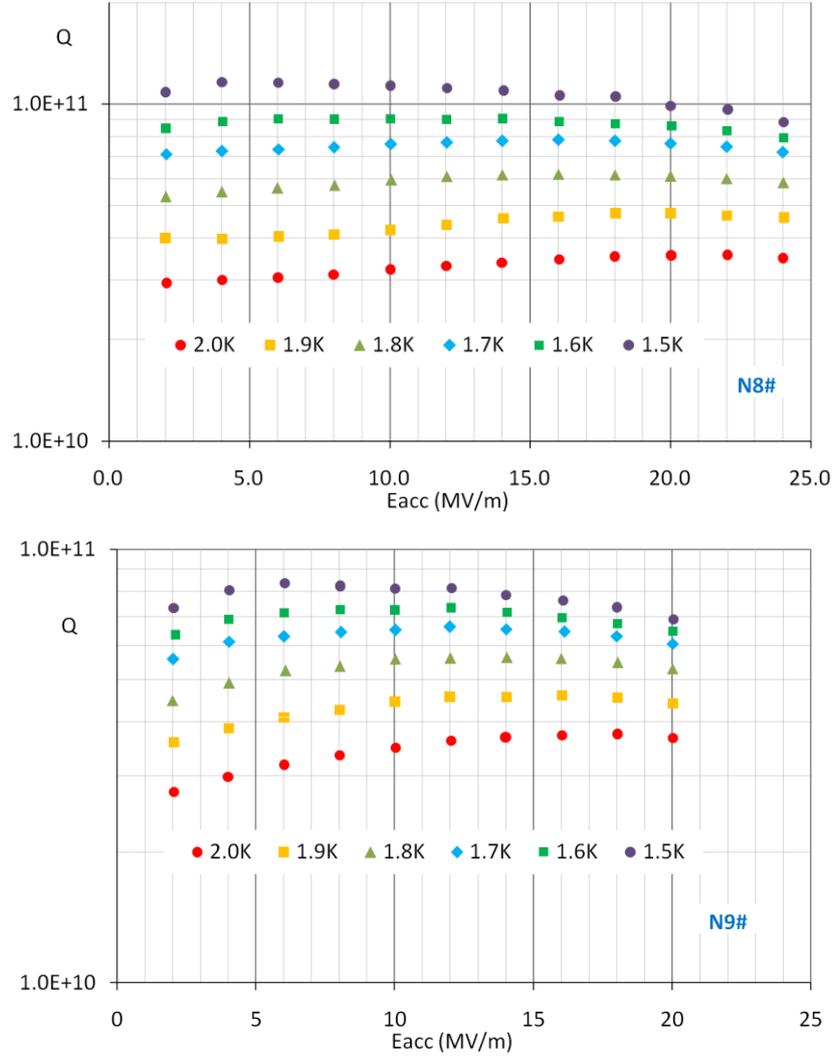

**Fig. 10** Quality factor versus gradient at 1.5K ~ 2.0K for N8# (up), N9# (down)

The surface resistance $R_S$ is composed of two parts: temperature-dependent BCS resistance ($R_{BCS}$) and residual resistance ($R_0$) [13].

$$R_S = R_{BCS} + R_0 \qquad (1)$$

$$R_{BCS} = A\frac{f^2}{T}\exp\left(-\frac{\Delta(T)}{kT}\right) \qquad (2)$$

Where $T$ is the temperature, $k$ is the Boltzmann constant, $2\Delta$ is the energy gap of superconductor, $f$ is the frequency of Superconducting cavity; $A$ is a constant which depends on superconducting parameters, such as coherence length, mean free path and London penetration depth.

According to the equation above, $R_{BCS}$ and $R_0$ have been calculated from the data from Fig. 11. The $R_{BCS}$ of both cavities at different temperatures is shown as Fig. 11, while $R_0$ is as Fig. 12. The $R_{BCS}$ decreases along with the increase of gradient between 5 and 20 MV/m, especially at 2.0 K. And the residual resistance increases with the gradient (see Fig. 12). So the anti-Q-slope behavior of cavity mid-T furnace baked is derived from the BCS resistance, which is similar as Nitrogen doping [14]. The residual resistance is 2~4 nΩ, which is higher than 0.63±0.06 nΩ at FNAL and ~1 nΩ at KEK.

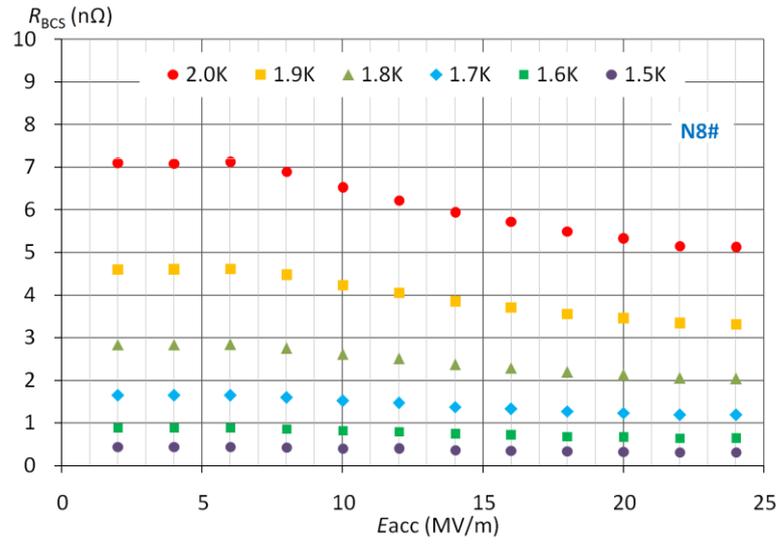

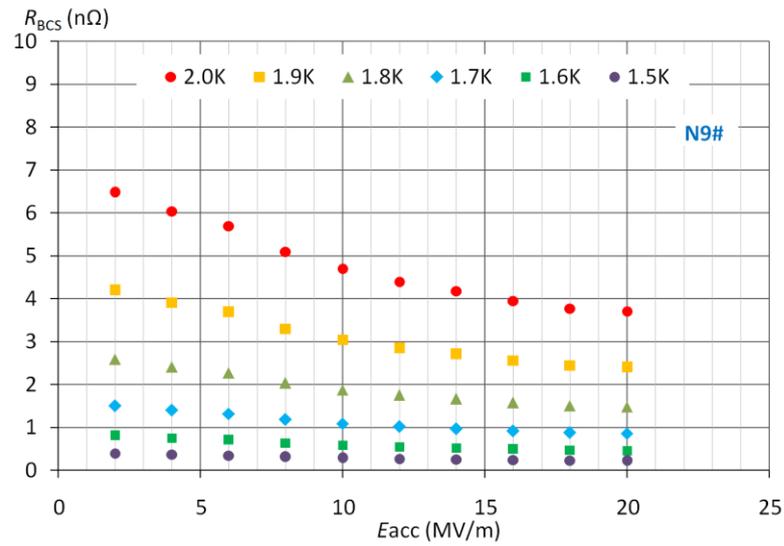

**Fig. 11** BCS resistance ($R_{BCS}$) of 1.3 GHz 9-cell cavities at different temperatures (up: N8#, down: N9#)

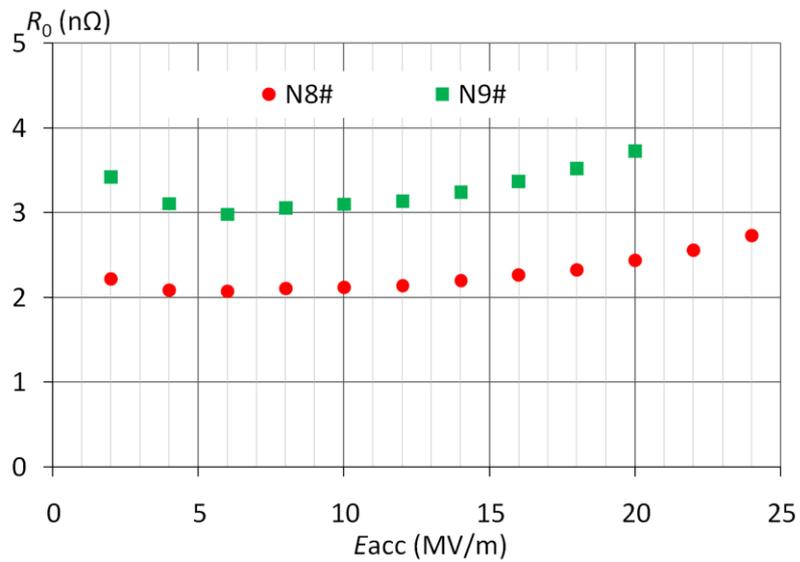

**Fig. 12** Residual resistance ($R_0$) of 1.3 GHz 9-cell cavities (N8#, N9#)

## 4. Conclusion

Study on the recipe of mid-T furnace bake has been carried out with 1.3 GHz 1-cell and 9-cell cavities, which have all shown improvement of Q value in the vertical test. Average Q of 9-cell cavities has reached $3.8\times10^{10}$ @ 16 MV/m, while the maximum gradients are between 22.7 and 26.5 MV/m. The behavior of anti-Q-slope during the vertical test is obvious. Besides N-doping, mid-T furnace bake may be another attractive candidate for high Q SRF accelerators.